\documentclass[superscriptaddress,nobibnotes,aps,prd,showkeys,showpacs,nofootinbib]{revtex4}
\usepackage{amsmath}

\usepackage{amsfonts,amssymb,amsmath,hyperref}
\usepackage{amssymb}
\usepackage{graphicx}
\usepackage{tikz}
\makeatletter
\newcommand*{\rom}[1]{\expandafter\@slowromancap\romannumeral #1@}
\makeatother
\usepackage{epsfig}

\begin{document}
%%%%%%%%%%%%%%%%%%%%%%%%%%%%%%%%%%%%%%%%%%%%%%%%%%%%%%%%%%%%%%%%%%%%%%%%%%%%%%%%

%%%%%%%%%%%%%%%%%%%%%%%%%%%%%%%%%%%%%%%%%%%%%%%%%%%%%%%%%%%%%%%%%%%%%%%%%%%%%%%%
\title{Fluid Dynamical Description of Sine-Gordon Solitons}
%%%%%%%%%%%%%%%%%%%%%%%%%%%%%%%%%%%%%%%%%%%%%%%%%%%%%%%%%%%%%%%%%%%%%%%%%%%%%%%%
\author{Nematollah Riazi}
\email{n\_riazi@sbu.ac.ir}
\affiliation{Physics Department, Shahid Beheshti University, Evin, Tehran 19839, Iran}
%%%%%%%%%%%%%%%%%%%%%%%%%%%%%%%%%%%%%%%%%%%%%%%%%%%%%%%%%%%
%\author{Maryam Riazi}
%\email{riazi_maryam@physics.sharif.edu}
%\affiliation{Physics Department, Sharif University of Technology, Tehran, Iran}
%%%%%%%%%%%%%%%%%%%%%%%%%%%%%%%%%%%%%%%%%%%%%%%%%%%%%%%%%%%%%%%%%%%%%%%%%%%%%%%%
\author{Marzieh Peyravi}
\email{marziyeh.peyravi@stu-mail.um.ac.ir}
\affiliation{Department of Physics, School of Sciences, Ferdowsi University
of Mashhad, Mashhad 91775-1436, Iran}
%%%%%%%%%%%%%%%%%%%%%%%%%%%%%%%%%%%%%%%%%%%%%%%%%%%%%%%%%%%%%%%%%%%%%%%%%%%%%%%%

%%%%%%%%%%%%%%%%%%%%%%%%%%%%%%%%%%%%%%%%%%%%%%%%%%%%%%%%%%%%%%%%%%%%%%%%%%%%%%%%
\begin{abstract}
We present a fluid dynamical description of a relativistic scalar field in $1+1$ dimensions and apply the general results to the special case of Sine-Gordon solitons. The results which include the local quantities pressure, density and fluid velocity are compared to the standard quantities attributed to the solitons.
\\
\end{abstract}

%%%%%%%%%%%%%%%%%%%%%%%%%%%%%%%%%%%%%%%%%%%%%%%%%%%%%%%%%%%%%%%%%%%%%%%%%%%%%%%%

\maketitle

%%%%%%%%%%%%%%%%%%%%%%%%%%%%%%%%%%%%%%%%%%%%%%%%%%%%%%%%%%%%%%%%%%%%%%%%%%%%%%%%
\section{Introduction}\label{Intr}
%%%%%%%%%%%%%%%%%%%%%%%%%%%%%%%%%%%%%%%%%%%%%%%%%%%%%%%%%%%%%%%%%%%%%%%%%%%%%%%%

The energy-momentum tensor is a second rank, symmetric and conserved tensor which is of central importance in fundamental relativistic theories like electromagnetism and general theory of relativity (see e.g. \cite{a,b}). For relativistic fields, it accommodates important local quantities like density, pressure, viscosity, fluid velocity and flux. Using the celebrated Noether's theorem, the energy-momentum tensor can be derived from the space and time
translation symmetries of the corresponding action or Lagrangian \cite{c,d}.
In general relativity, the energy-momentum tensor which is the source of space-time curvature is calculated by varying the matter action ($\cal{A}$) with respect to the metric \cite{Wein}:
\begin{equation}\label{1}
T^{\mu\nu}=-\frac{2}{\sqrt{-g}}\frac{\delta \mathcal{A}_{m}}{\delta g_{\mu\nu}},
\end{equation}
where $T^{\mu\nu}$ is the energy-momentum tensor, $g_{\mu\nu}$\footnote{We adopt the metric signature
$(-,+,+,+)$} is the metric tensor, $\mathcal{L}_{m}$ is the matter Lagrangian, and $g$ is the determinant of the metric tensor. In flat space-time, the energy-momentum conservation takes the simple form \cite{RD}:
\begin{equation}\label{2}
\partial_{\mu}T^{\mu\nu}=0.
\end{equation}
where sum over repeated indices is implied. For $\nu=0$, we can expand Eq. (\ref{2}) as \cite{RD}:
\begin{equation}\label{3}
\partial_{0}T^{00}+\partial_{i}T^{i0}=0.
\end{equation}
which is the continuity equation $\frac{\partial\rho}{\partial t}+\nabla\cdot\vec{j}=0$, with $\rho\equiv T^{00}$ and $j_{i}\equiv T^{i0}=T^{0i}$.
Integrating Eq. (\ref{3}) over the whole Euclidean space, doing an integration by part and assuming that $\vec{j}$ vanishes over the boundary, we obtain \cite{RD}:
\begin{equation}\label{4}
\frac{d}{dt}\int_{V} T^{00} d^{3}x+\oint\vec{j}\cdot d\vec{s}=0,
\end{equation}
or $\frac{dE}{dt}=0$, where $E=\int T^{00} d^{3}x$ is the total conserved energy. It is interesting that for a single, moving soliton (kink), Eq. (\ref{3}) leads to the famous Einstein equation \cite{RD}:
\begin{equation}\label{emc}
E^2=Mc^2+p^2c^2,
\end{equation}
in which $M$ and $p$ are the total rest energy (mass) and momentum of the soliton,
respectively \cite{e}. 

On the other hand, the relativistic energy-momentum tensor of an imperfect, relativistic fluid is given by \cite{6,RD}:
\begin{equation}\label{5}
T^{\mu\nu}=\rho u^{\mu}u^{\nu}+q^{(\mu}u^{\nu)}+S^{\mu\nu},
\end{equation}
where $\rho$ is the co-moving energy density, $ u^{\mu}$ is the fluid 4-velocity, and $q^{\mu}$ is the space-like heat vector orthogonal to $u^{\mu}$ (i.e. $q^{\mu}u_{\mu}=0$). For sub-luminal fluid  velocity $u^{\alpha}u_{\alpha}=-1$, while luminal and super-luminal motion correspond to $u^{\alpha}u_{\alpha}=0$ and $u^{\alpha}u_{\alpha}=+1$, respectively. $S^{\mu\nu}$ is the stress tensor with eigenvectors defining the directions of the principal stresses.

In the Eckhart formulation \cite{7,8}, the energy-momentum tensor is written in terms of viscosity as:
\begin{equation}\label{6}
T^{\mu\nu}=\rho u^{\mu}u^{\nu}+q^{(\mu}u^{\nu)}+ \left(p-J\Theta\right)h^{\mu\nu}-2\eta\sigma^{\mu\nu}.
\end{equation}
where the positive quantities $J$ and $\eta$ are the bulk and shear viscosities, respectively, $\Theta\equiv\nabla_{\mu}u^{\mu}$ is the expansion,
$h^{\mu\nu}\equiv\eta^{\mu\nu}+u^{\mu}u^{\nu}$, and $\sigma^{\mu\nu}$ is the symmetric, traceless, shear tensor:
\begin{equation}\label{7}
\sigma^{\mu\nu}\equiv\frac{1}{2}\left(\nabla_{\alpha}u^{\mu}\eta^{\alpha\nu}+\nabla_{\alpha}u^{\nu}\eta^{\alpha\mu}\right)-\frac{1}{3}\Theta\eta^{\mu\nu}.
\end{equation}
For a perfect fluid in which the velocity gradients and heat flux can be ignored, $\Theta=0$, $\sigma^{\mu\nu}=0$, $q^{\mu}=0$ and we have:
\begin{equation}\label{8}
T^{\mu\nu}= \left(\rho+p\right)u^{\mu}u^{\nu}+p\eta^{\mu\nu}.
\end{equation}
The canonical theory for perfect fluids, in Eulerian and Lagrangian formulations in relation to a description of extended structures in higher dimensions is reviewed in \cite{11}.

On the other hand, if the system under consideration is composed of a statistically large number of relativistic particles of the rest mass $m$ and  4-momenta $p^{\mu}$, the energy-momentum tensor can be calculated via \cite{b,8a}:
\begin{equation}\label{8a}
T^{\mu\nu}=\int\frac{f(\vec{x},\vec{p})p^{\mu}p^{\nu}d^{3}p}{p^{0}},
\end{equation}
in which $f$ is the number of particles per unit volume of phase space ($\vec{x},\vec{p}$), and $p^{\mu}p_{\mu}=-m^{2}$ with $m$ being the rest mass of constituent particles ($\hbar=c=1$ is assumed).

 It is well-known that solitons have a dual identity at the same time, corresponding to particle-like and wave-like properties. By considering solitons as particles in microscopic point of view , statistical mechanics will be an effective and powerful method for studying their interactions. However, their macroscopic behaviour can be represented in terms of the hydrodynamics of soliton gas which can be interpreted as an integrable turbulence \cite{Dh, TB, zak, EL, BD}. Generalised hydrodynamics of soliton gases has been introduced and studied for KdV solitons recently \cite{Dh,TB, EL, BD}. 

In this context, our main question, to be answered in this paper is the following: what are the relationships between the field quantities like $\rho$ and $w$ (soliton velocity) and those appearing the relativistic fluid description ($\rho$, $p$, $u^{\alpha}$), if the soliton is to be described by a perfect fluid. Besides some expected results, there appear new interesting relations, particularly in relation to the concept of soliton velocity, and the possibility of super-lunimal motion.

In the next section, we apply the above-mentioned concepts and definitions to the case of a relativistic scalar field in $1+1$ dimension, and will then concentrate on the special case of Sine-Gordon solitons.

%%%%%%%%%%%%%%%%%%%%%%%%%%%%%%%%%%%%%%%%%%%%%%%%%%%%%%%%%%%%%%%%%%%%%%%%%%%%%%%%
\section{ Scalar Fields in $1+1$ Dimensions}\label{sec2}
%%%%%%%%%%%%%%%%%%%%%%%%%%%%%%%%%%%%%%%%%%%%%%%%%%%%%%%%%%%%%%%%%%%%%%%%%%%%%%%%

The action of a real scalar field in $1+1$ dimensions is given by \cite{d}:
\begin{equation}\label{9}
\mathcal{A}=\int d^{2}x \left(-\frac{1}{2}\partial^{\mu}\varphi\partial_{\mu}\varphi-V(\varphi)\right).
\end{equation}
where $V(\varphi)$ is the self-interaction potential. Variation of  Eq. (\ref{9}) with respect to $\varphi$ leads to the equation of motion \cite{d}:
\begin{equation}\label{10}
\square \varphi=\partial^{\mu}\partial_{\mu}\varphi=\frac{\partial V}{\partial \varphi},
\end{equation}
while the energy-momentum tensor is given by \cite{RD}:
\begin{equation}\label{11}
T^{\mu\nu}=\partial^{\mu}\varphi\partial^{\nu}\varphi+\eta^{\mu\nu}\mathcal{L},
\end{equation}
$\mathcal{L}$ being the Lagrangian density.
We will equate this to the energy-momentum tensor of a perfect fluid Eq. (\ref{8}), as far as all the results for $\rho$, $p$ and $u^{\alpha}$ are mutually consistent. In the case where inconsistencies arise, we may conclude that the scalar field cannot be described by a perfect fluid and we should resort to a more complicated fluid model. For the time being, we keep the three possibilities $u^{\alpha}u_{\alpha}=\varepsilon$, with $\varepsilon=-1,0,+1$ which correspond to time-like, light-like and space-like or super-luminal motion of the fluid, respectively.

Let us start with the simple case of a massless scalar field with $V(\varphi)=0$. The scalar field then obeys a wave equation and any function $\varphi(x \pm t)$
will be a solution ($c=1$ is assumed). Assuming the wave is moving in the $+x$ direction, we can use the light-cone variable
$\xi\equiv x-t$. We have:
\begin{equation}\label{12}
T^{00}=(\rho+p)(u^{0})^{2}-p=\left(\frac{\partial\varphi}{\partial t}\right)^2-\mathcal{L}.
\end{equation}
where $\mathcal{L}=\frac{1}{2}\dot{\varphi}^{2}-\frac{1}{2}\varphi'^{2}$, dot and prime being derivatives with respect to $t$ and $x$, respectively. We have:
\begin{eqnarray}\label{13}
\label{true1}\hspace{1cm}\dot{\varphi}&=&\frac{\partial\varphi}{\partial t}=\frac{\partial\xi}{\partial t}\frac{\partial\varphi}{\partial\xi}=-\varphi_{\xi},\nonumber\\
\label{true2}\hspace{1cm}\varphi'&=&\frac{\partial\varphi}{\partial x}=\frac{\partial\xi}{\partial x}\frac{\partial\varphi}{\partial\xi}=\varphi_{\xi}.
\end{eqnarray}
The wave equation $\Box \varphi=0$ is trivially satisfied and we have $\mathcal{L}=0$ identically. For the energy density we obtain:
\begin{equation}\label{14}
T^{00}=\left(\frac{\partial\varphi}{\partial t}\right)^{2}=\frac{1}{2}\left(\frac{\partial\varphi}{\partial t}\right)^{2}+\frac{1}{2}\left(\frac{\partial\varphi}{\partial t}\right)^{2}=\frac{1}{2}\dot{\varphi}^2+\frac{1}{2}\varphi'^2=\varphi_{\xi}^{2},
\end{equation}
the last step being made using $\varphi'=-\dot{\varphi}$ from Eq. (\ref{13}).
Let $u^{\alpha}\equiv (\gamma u,\gamma v)$ with $u^{\alpha}u_{\alpha}=\varepsilon=-1,0,+1$. We thus obtain:
\begin{eqnarray}\label{1517}
\label{true3}\hspace{1cm}T^{00}&=&\varphi_{\xi}^{2}=\left(\rho+p\right)\gamma^2 u^{2}-p,\\
\label{true4}\hspace{1cm}T^{11}&=&\left(\frac{\partial \varphi}{\partial x}\right)^{2}=\varphi_{\xi}^{2}=\left(\rho+p\right)\gamma^2 v^{2}+p,\\
\label{true5}\hspace{1cm}T^{01}&=&\partial^{0}\varphi\partial^{1}\varphi=-\frac{\partial\varphi}{\partial t}\frac{\partial\varphi}{\partial x}=\varphi_{\xi}^{2}=\left(\rho+p\right)\gamma^2 uv,
\end{eqnarray}
or
\begin{equation}\label{18}
T^{\mu\nu}=\left(%
\begin{array}{cc}
  1 & 1 \\
  1 & 1 \\
\end{array}%
\right)\varphi_{\xi}^{2}.
\end{equation}
Note that there is no spatial shear in $1+1$ dimensions and $T^{\mu}_{\mu}=0$ with $V=0$ which is also implied by Eq. (\ref{18}). We have:
\begin{equation}
T^{\mu}_{\mu}=\left(\rho+p\right)u^{\mu}u_{\mu}+2p=\left(\rho+p\right)\varepsilon+2p=\begin{cases}
2p    &   \textrm{($u^{\mu}u_{\mu}=0$)}\\
  \rho+3p &  \textrm{($u^{\mu}u_{\mu}=+1$)}\\
  -\rho+p &  \textrm{($u^{\mu}u_{\mu}=-1$)},\\
\end{cases}
\end{equation}
while Eq. (\ref{11}) leads to:
\begin{equation}
T^{\mu}_{\mu}=\partial^{\mu}\varphi\partial_{\mu}\varphi+2\left(-\frac{1}{2}\partial^{\mu}\varphi\partial_{\mu}\varphi\right)=0,
\end{equation}
We therefore obtain:
\begin{equation}
p=-\frac{\epsilon}{\epsilon+2}\rho.
\end{equation}

Equations (\ref{true3})-(\ref{true5}) constitute three equation for four unknowns $\rho$, $p$, $u$, $v$ once $\varphi(x-t)$ is known as a uniformly moving solution of the wave equation. The remaining equation comes from one of the conditions $\varepsilon\equiv u^{\mu}u_{\mu}=\gamma^2(-u^2+v^2)=0,\pm1$. 

For the general case of a scalar field with potential ($V(\varphi)\neq0$), we rewrite our fundamental equations in the following form:
\begin{eqnarray}
\label{true18}\hspace{1cm}\pi \gamma^2 u^{2}-p&=&K+V,\\
\label{true19}\hspace{1cm}\pi \gamma^2 v^{2}+p&=&K-V,\\
\label{true20}\hspace{1cm}\pi \gamma^2 uv&=&-\psi,\\
\label{true21}\hspace{1cm}u^{\mu}u_{\mu}&=&\gamma^2(v^{2}-u^{2})=\varepsilon,
\end{eqnarray}
where $K\equiv\frac{1}{2}\varphi'^{2}+\frac{1}{2}\dot{\varphi}^{2}$, $\pi\equiv p+\rho$ and $\varepsilon=0,\pm 1$. We also have $\mathcal{L}=-S-V$, where $S\equiv\frac{1}{2}\varphi'^{2}-\frac{1}{2}\dot{\varphi}^{2}$, $\psi\equiv\dot{\varphi}\varphi'$. From Eqs. (\ref{true18}), (\ref{true19}) and  (\ref{true21}) we obtain:
\begin{equation}\label{55}
\pi\left(2\gamma^2u^{2}+\varepsilon\right)=2K,
\end{equation}
and since $\pi^{2}\gamma^4u^{2}v^{2}=\psi^{2}$, we obtain
\begin{equation}\label{56}
\frac{\psi^{2}}{u^{2}\left(u^{2}+\varepsilon/\gamma^2\right)}\left(2u^{2}+\varepsilon/\gamma^2\right)^{2}=4K^{2},
\end{equation}
Eq. (\ref{true18}) minus Eq. (\ref{true19}) gives:
\begin{equation}\label{57}
-\pi\varepsilon-2p=2V,
\end{equation}
solving Eq. (\ref{56}) for $u^{2}$ gives:
\begin{equation}\label{58}
\gamma^2 u^{2}=-\frac{\epsilon}{2}\pm\frac{1}{\sqrt{1-\psi^2/K^2}}.
\end{equation}

Once $\gamma^2 u^{2}(x,t)$ is obtained using Eq. (\ref{58}) in terms of the solution $\varphi(x,t)$, $\gamma^2\pi=\frac{2K}{2u^{2}+\varepsilon}$ or $\pi=-\frac{\psi}{u\sqrt{u^{2}+\varepsilon}}$ is immediately calculated, from which $p$ is obtained
using $p=K-V-\pi\left(u^{2}+\varepsilon\right)$, $\rho$ from $\rho=\pi-p$ and $v^{2}=u^{2}-\varepsilon$.
For the null case, equations simplify considerably since $\varepsilon=0$ and  $u^{2}=v^{2}$.

%%%%%%%%%%%%%%%%%%%%%%%%%%%%%%%%%%%%%%%%%%%%%%%%%%%%%%%%%%%%%%%%%%%%%%%%%%%%%%%%
\section{Hydrodynamics of Sine-Gordon Solitons}\label{sec3}
%%%%%%%%%%%%%%%%%%%%%%%%%%%%%%%%%%%%%%%%%%%%%%%%%%%%%%%%%%%%%%%%%%%%%%%%%%%%%%%%

\begin{figure}
\epsfxsize=12cm\centerline{\epsfbox{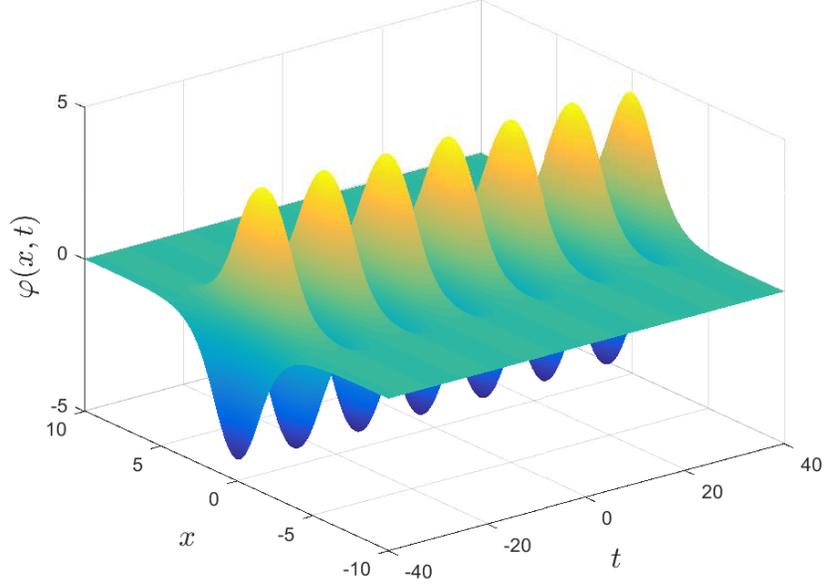}}\caption{The breather solution of the SG system during $-40\leq t\leq 40$ and in the interval $-10<x<10$.\label{fig:br}}
\end{figure}

We are now in a position to do the same calculations for the interesting case of the well-known Sine-Gordon solitons. The Sine-Gordon system is the prototype
of a relativistic integrable system. The self-interaction potential of this system in dimensionless units is given by \cite{c,e,10, man, Raj, Vach, Ryder, horita}:
\begin{equation}\label{37}
V(\varphi)=1-\cos(\varphi),
\end{equation}
leading to the field equation:
\begin{equation}\label{38}
\Box\varphi=\sin(\varphi).
\end{equation}
Single soliton solutions of this equation which are called kinks (and anti-kinks) are given by:
\begin{equation}\label{39}
\varphi(x,t)=\pm4\arctan\Big(\exp\left(\gamma_w\left(x-wt\right)\right)\Big),
\end{equation}
where $w$ is the soliton velocity and $\gamma_w\equiv\left(1-w^{2}\right)^{-1/2}$. Multi-soliton solutions of the SG equation can be derived using various methods like inverse scattering  \cite{10} or B\"{a}cklund transformations \cite{c}. Besides the single soliton (kink) solution Eq. (\ref{39}), we concentrate only on the kink-antikink and breather solutions \cite{c,e};\\
kink-antikink \cite{9}:
\begin{equation}\label{40}
\varphi(x,t)=4\arctan\left[\frac{\sinh\left(\gamma_w wt\right)}{w\cosh\left(\gamma_w x\right)}\right],
\end{equation}
and breather \cite{9}:
\begin{equation}\label{41}
\varphi(x,t)=4\arctan\left[\frac{\frac{\sin(wt)}{\sqrt{1+w^{2}}}}{\frac{w\cosh(x)}{\sqrt{1+w^{2}}}}\right].
\end{equation}
As can be seen from the figures, the first solution (Fig. \ref{fig:kak}), describes an anti-kink moving from left and a kink moving from right toward each other, interacting around $x\sim 0$ and $t\sim 0$, and then moving apart with the same velocities. The breather solution (Fig. \ref{fig:br}) describes a bound state of kink-antikink, swinging around their ``center of mass'' indefinitely. 

%%%%%%%%%%%%%%%%%%%%%%%%%%%%%%%%%%%%%%%%%
\subsection{Uniformly Moving Kink} \label{sec31}
%%%%%%%%%%%%%%%%%%%%%%%%%%%%%%%%%%%%%%%%%

For the uniformly moving kink solution Eq. (\ref{39}), we obtain:
\begin{equation}
\frac{K}{\psi}=-\frac{1+w^2}{2w},
\end{equation}
 $\gamma^2u^2=1/(1-w^2)=\gamma_w^2$ for the + solution and $\epsilon=-1$. We then have $\gamma^2v^2=w^2 \gamma_w^2$. It is seen that we are led to the expected result $u=1$ and $v=w$, only if we adopt $\epsilon=-1$ (time-like solution), and select the + solution. In what follows, we keep using these choices for two-soliton solution (kink-antikink and kink-kink). 

%%%%%%%%%%%%%%%%%%%%%%%%%%%%%%%%%%%%%%%%%%%%%%%%%%%%%%%%%%%%%%%%%%%%%%%%%%%%%%%%
\subsection{Two-soliton Solutions} \label{sec4}
%%%%%%%%%%%%%%%%%%%%%%%%%%%%%%%%%%%%%%%%%%%%%%%%%%%%%%%%%%%%%%%%%%%%%%%%%%%%%%%%

\begin{figure}
\epsfxsize=12cm\centerline{\epsfbox{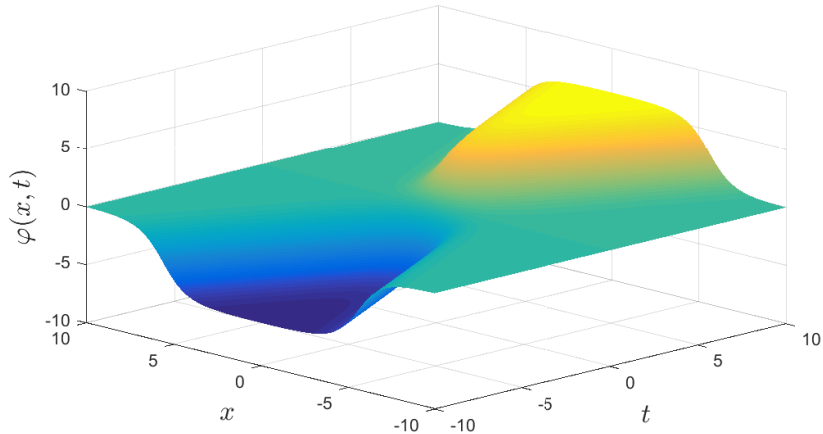}}\caption{The kink-antikink solution of the SG system. The two oppositely charged solitons
interact non-linearly around $t\approx0$ and $x\approx0$, and then leave the interaction region without any change in shape. \label{fig:kak}}
\end{figure}

The exact solution Eq. (\ref{40}) describes the scattering of a kink with an anti-kink, with opposite topological charges\footnote{ Topological current is defined by $J^{\mu}=\frac{1}{2\pi}\partial^{\mu}\varphi$, from which $Q=\int_{-\infty}^{+\infty}J^{0}dx=\frac{1}{2\pi}\left[\phi(+\infty)-\phi(-\infty)\right].$}
and interaction region around $t\approx0$ and $x\approx0$ (see Fig. \ref{fig:kak}).
%%%%%%%
%%%%%%%

%%%%%%%
%%%%%%%
Both of the two-soliton solutions Eqs. (\ref{40}) and (\ref{41}) are in the following form:
\begin{equation}
\varphi(x,t)=4\arctan\left(\frac{f(t)}{g(x)}\right).
\end{equation}
which implies:
\begin{eqnarray}
\label{true31}\hspace{1cm}\dot{\varphi}&=&\frac{4g\dot{f}}{f^{2}+g^{2}},\\
\label{true32}\hspace{1cm}\varphi'&=&-\frac{4fg'}{f^{2}+g^{2}},\\
\label{true33}\hspace{1cm}K&=&\frac{8\left(g^{2}\dot{f}^{2}+f^{2}g'^{2}\right)}{\left(f^{2}+g^{2}\right)^{2}},\\
\label{true332}\hspace{1cm}\psi&=&-\frac{16fg\dot{f}g'}{(f^2+g^2)^2},\\
\label{true34}\hspace{1cm}S&=&\frac{8\left(f^{2}g'^{2}-g^{2}\dot{f}^{2}\right)}{\left(f^{2}+g^{2}\right)^{2}},\\
\label{true35}\hspace{1cm}\pi&=&2S=\frac{16\left(f^{2}g'^{2}-g^{2}\dot{f}^{2}\right)}{\left(f^{2}+g^{2}\right)^{2}},\\
\label{true36}\hspace{1cm}p&=&S-V,\\
\label{true37}\hspace{1cm}\rho&=&S+V,\\
\label{true38}\hspace{1cm}\gamma^2u^2&=&\frac{1}{2}+\frac{g^2\dot{f}^2+f^2(g')^2}{|g^2\dot{f}^2-f^2(g')^2|},\\
\label{true39}\hspace{1cm}\gamma^2v^2&=&\frac{g^2\dot{f}^2+f^2(g')^2}{|g^2\dot{f}^2-f^2(g')^2|}.
\end{eqnarray}
As a check for the correctness of these relations, we apply them to the case of moving kink with $\varphi(x,t)=4\arctan\left(\exp(\gamma(x-wt))\right)$, for which $f(t)=\exp(-\gamma wt)$, $g(x)=\exp(-\gamma x)$, $\gamma_w=\left(1-w^{2}\right)^{-1/2}$, $w$ being the kink velocity ($|w|<1$). Straightforward calculation yields $\dot{f}=-\gamma wf$, $g'=-\gamma g$
\begin{eqnarray}
\label{true40}\hspace{1cm}S&=&\frac{8f^{2}g^{2}}{\left(f^{2}+g^{2}\right)^{2}}=\frac{\pi}{2},\\
\label{true41}\hspace{1cm}p&=&\frac{8f^{2}g^{2}}{\left(f^{2}+g^{2}\right)^{2}}-V=S-V=\left(\frac{1-w^{2}}{1+w^{2}}\right)K-V,\\
\label{true42}\hspace{1cm}\rho&=&\frac{8f^{2}g^{2}}{\left(f^{2}+g^{2}\right)^{2}}+V=S+V=\left(\frac{1-w^{2}}{1+w^{2}}\right)K+V,\\
\label{true43}\hspace{1cm}\gamma^2u^{2}&=&\frac{1}{1-w^{2}}=\gamma_w^{2},\\
\label{true44}\hspace{1cm}\gamma^2v^{2}&=&\gamma_w^{2}-\varepsilon=\gamma_w^{2}w^{2},   \qquad  (\varepsilon=-1)\\
\end{eqnarray}

Note that $u^{\alpha}=\left(\gamma,\gamma w\right)$ as expected, only if $\varepsilon=-1$. For the static kink ($w=0$, $\gamma=1$), we recover the expected results $\rho=K+V$ and $p=K-V$.

We can now get back to the general results Eqs. (\ref{true34})-(\ref{true39}), set $\varepsilon=-1$, and insert appropriate function for $f(t)$ and $g(x)$ for kink-antikink and breather solutions.\\
Kink-antikink scattering:
\begin{eqnarray}
f(t)&=&\sinh\left(\gamma_w wt\right),\nonumber\\
g(x)&=&w\cosh\left(\gamma_w x\right).
\end{eqnarray}
Breather:
\begin{eqnarray}
f(t)&=&\frac{\sin(wt)}{\sqrt{1+w^{2}}},\nonumber\\
g(x)&=&\frac{w\cosh(x)}{\sqrt{1+w^{2}}}.
\end{eqnarray}

\begin{figure}
\epsfxsize=12cm\centerline{\epsfbox{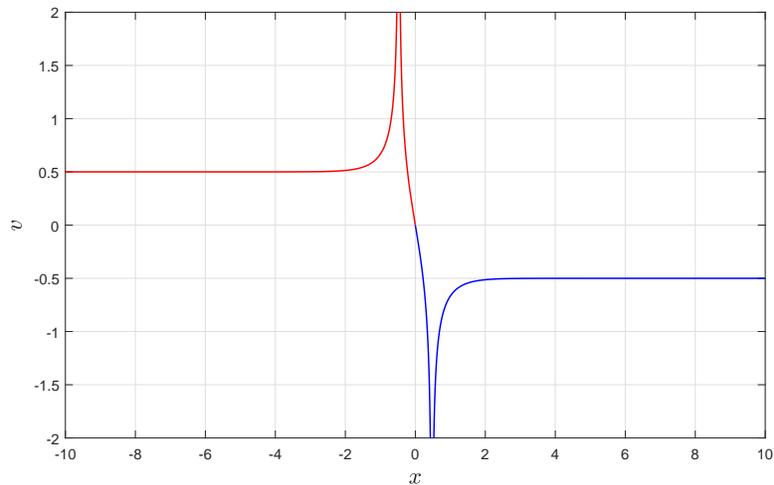}}
\caption{\label{fig:6} The fluid velocity $v$, as a function of $x$ for $t=-10$, before collision. The kink is moving with the velocity $w=0.5$, while the anti-kink is moving with velocity $w=-0.5$, toward each other. The interaction region is $x\sim 0$, around which the fluid velocity becomes largely different from the soliton velocity, and eventually becomes super-luminal.}
\end{figure}

\begin{figure}
\epsfxsize=12cm\centerline{\epsfbox{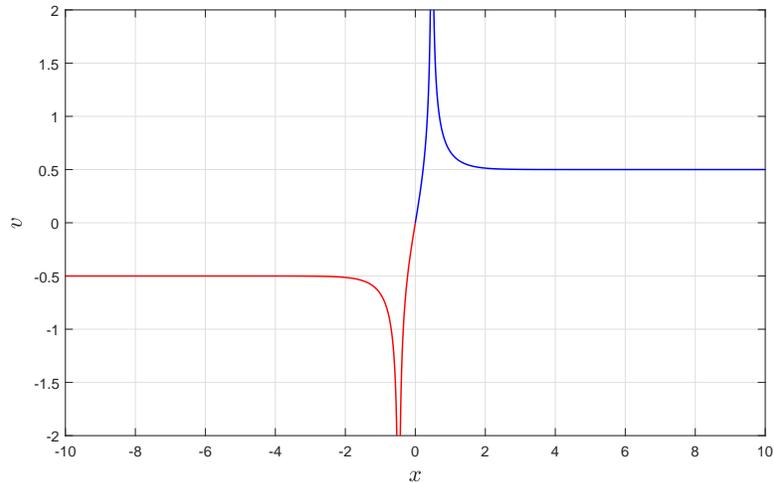}}
\caption{\label{fig:7} The fluid velocity $v$, as a function of $x$ for $t=10$, after collision. The kink is moving with the velocity $w=0.5$, while the anti-kink is moving with velocity $w=-0.5$, receding  each other. The interaction region is $x\sim 0$, around which the fluid velocity becomes largely different from the free soliton velocity.}
\end{figure}

\begin{figure}
\epsfxsize=12cm\centerline{\epsfbox{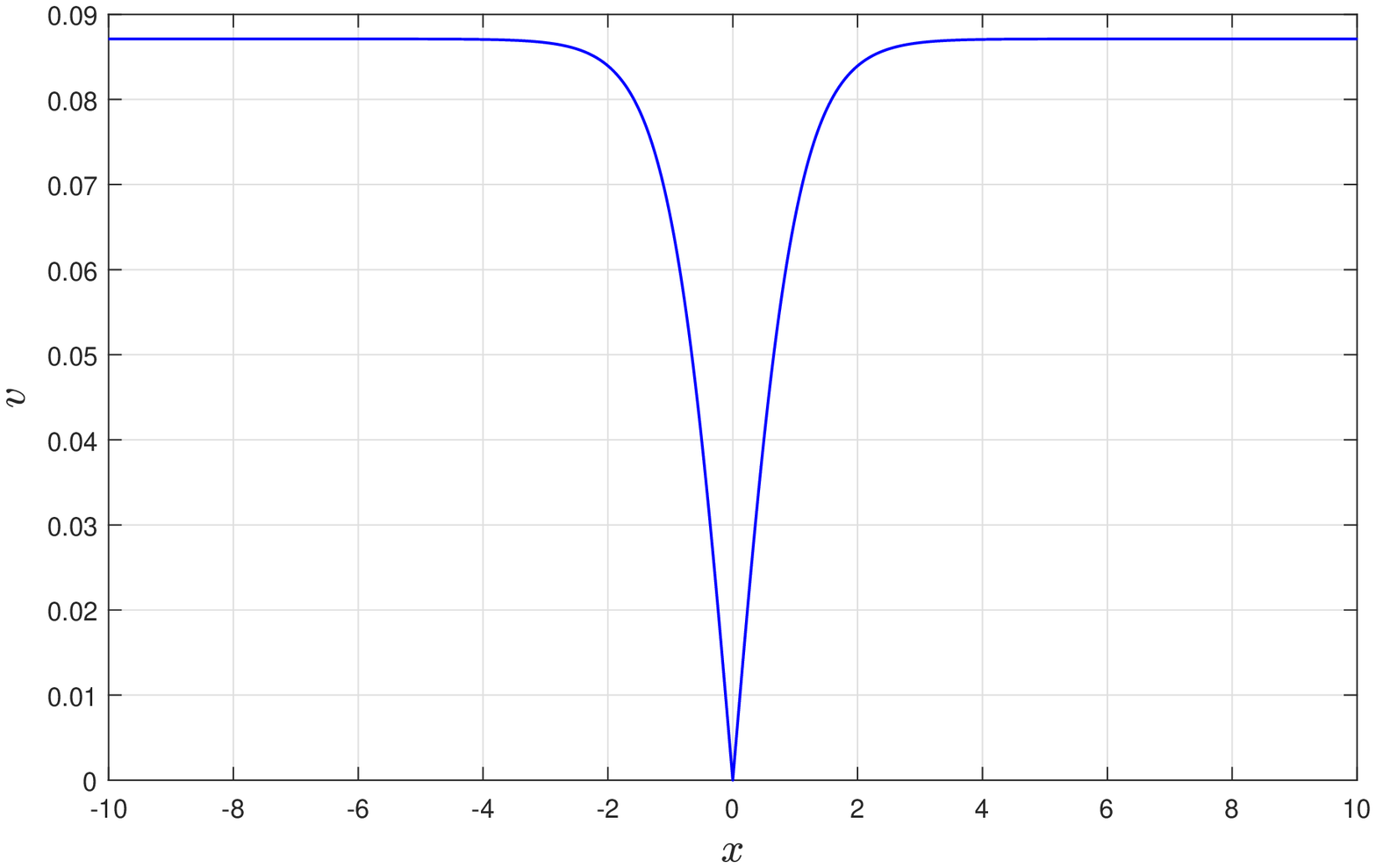}}
\caption{\label{fig:8} The fluid velocity $v$ of breather, as a function of $x$ for $t=0.1$ and $w=0.5$. }
\end{figure}

\begin{figure}
\epsfxsize=12cm\centerline{\epsfbox{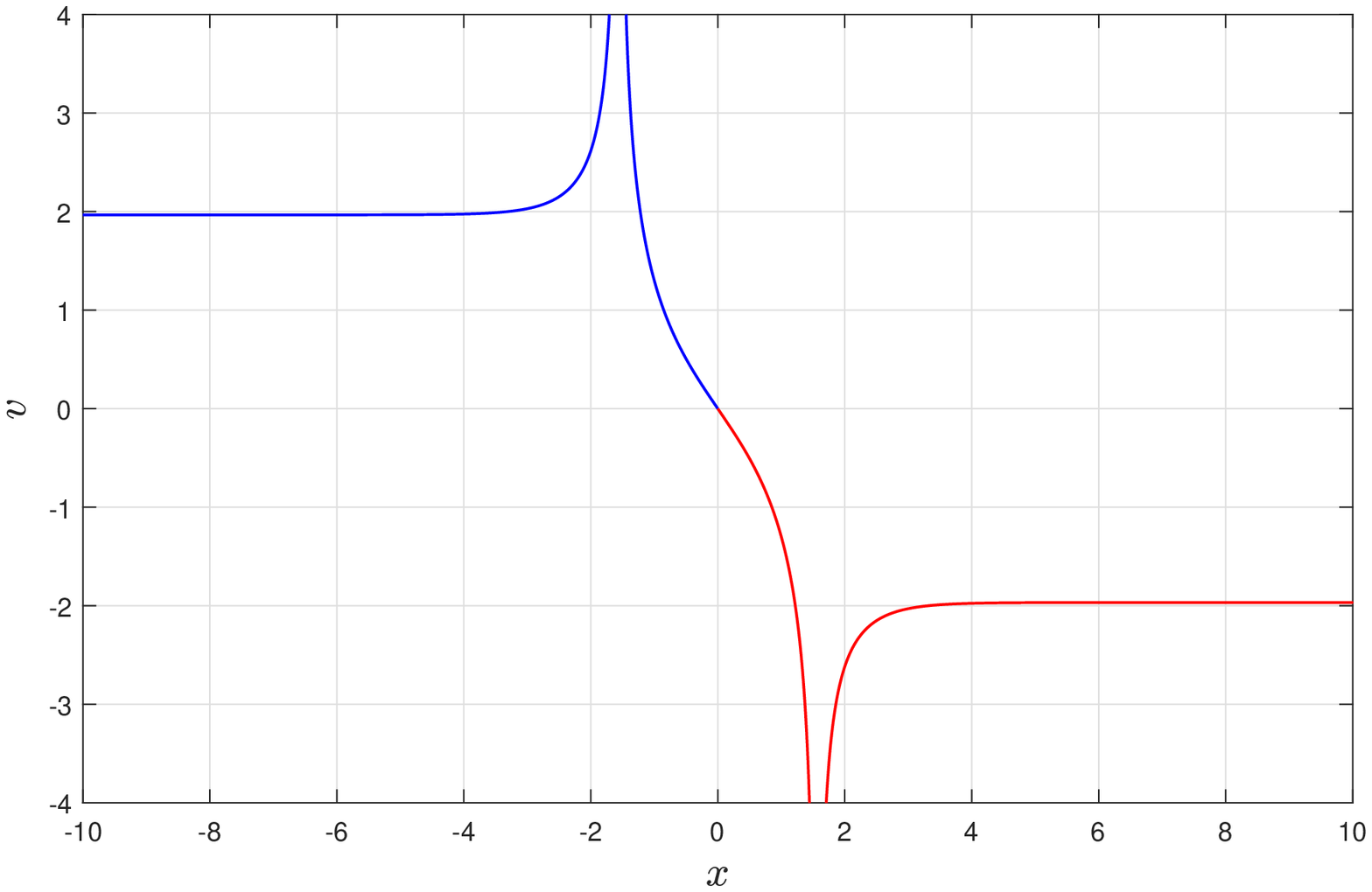}}
\caption{\label{fig:9} The fluid velocity $v$ of breather, as a function of $x$ for $t=1$ and $w=0.5$. }
\end{figure}

It is now easy to insert the expressions for $K$ and $\psi$ for each case and calculate the corresponding fluid velocity. For the scattering of a kink and an antikink, two solitons approach each other during $t \in (-\infty,0)$, interact around $t\sim 0$, and scatter off each other with almost constant velocity during $t \in (0,+\infty)$. For the asymptotic situation $t,x\rightarrow \infty$, we can use the approximation $\sinh\gamma x\simeq 1$ and $\sinh(\gamma wt)\sim 1$, which neatly leads to $v\simeq w$. This shows that the fluid velocity coincides with the collective soliton velocity as long as the solitons are far from the interaction region. The interesting question is: what is the fluid velocity in the interaction region? In order to answer this question, we plot the velocity field of the fluid $v$ as a function of $x$ for two values of $t$ in Figs. \ref{fig:6} and \ref{fig:7} for kink-antikink scattering, and Figs. \ref{fig:8} and \ref{fig:9} for the breather solution.  Straightforward calculation leads to:
\begin{equation}\label{am}
\gamma^2v^2=-\frac{1}{2}+\frac{1}{2}\frac{w^2+\tanh^2(\gamma_w x)\tanh^2(\gamma_w wt)}{\left|w^2-\tanh^2(\gamma_w x)\tanh^2\gamma_w (wt)\right|},
\end{equation}
for the kink-antikink scattering, and 
\begin{equation}\label{vanti}
\gamma^2v^2=-\frac{1}{2}+\frac{1}{2}\frac{w^2+\tan^2(wt)\tanh^2(x)}{\left|w^2-\tan^2(wt)\tanh^2(x)\right|}
\end{equation}
for the breather solution. It is also interesting to find the asymptotic value of the soliton velocity $v$, far from the interaction region. Putting $\tanh(\gamma x)$ and $\tanh(\gamma w t)$ in Eq. (\ref{am}) equal to one, we obtain $v=\pm w$, which correspond to two solitons approaching (or receding) each other from infinity, with velocities $\pm w$. Of course, the velocities do not remain constant as the solitons approach each other.  

\begin{figure}
\epsfxsize=12cm\centerline{\epsfbox{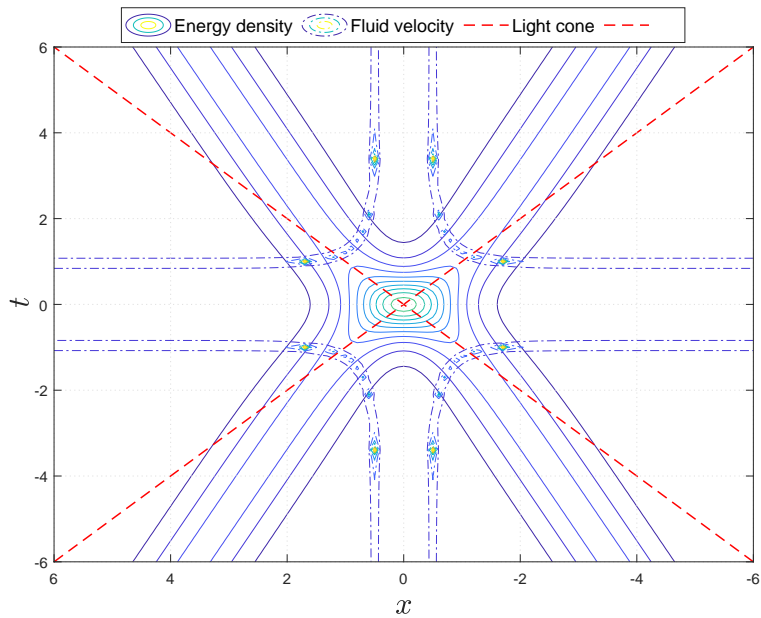}}
\caption{\label{fig:10} The fluid velocity ($v$) and  the energy density ($\rho$) peak for kink-antikink scattering versus the light cone in $t-x$ plane. }
\end{figure}

If we define the soliton velocity as the velocity of the energy density peak, we clearly see from Fig. \ref{fig:10} that it is nearly constant at large distances, approaches the velocity of light and eventually exceeds it near the center of mass!  There is nothing to worry about, as this velocity is not necessarily equal to the group velocity which should not be greater than $c=1$. Also drawn on the same figure, is the fluid velocity contours, which display the same behaviour. The fluid velocity defined via Eq. (\ref{vanti}), is constant at large distances and diverges in the vicinity of the interaction point. This observation, leads to the conclusion that the kink and antikink become space-like (cross the light cone) very near to the interaction region. 

%%%%%%%
%%%%%%%

\begin{figure}[h]
\epsfxsize=9cm\centerline{\hspace{8cm}\epsfbox{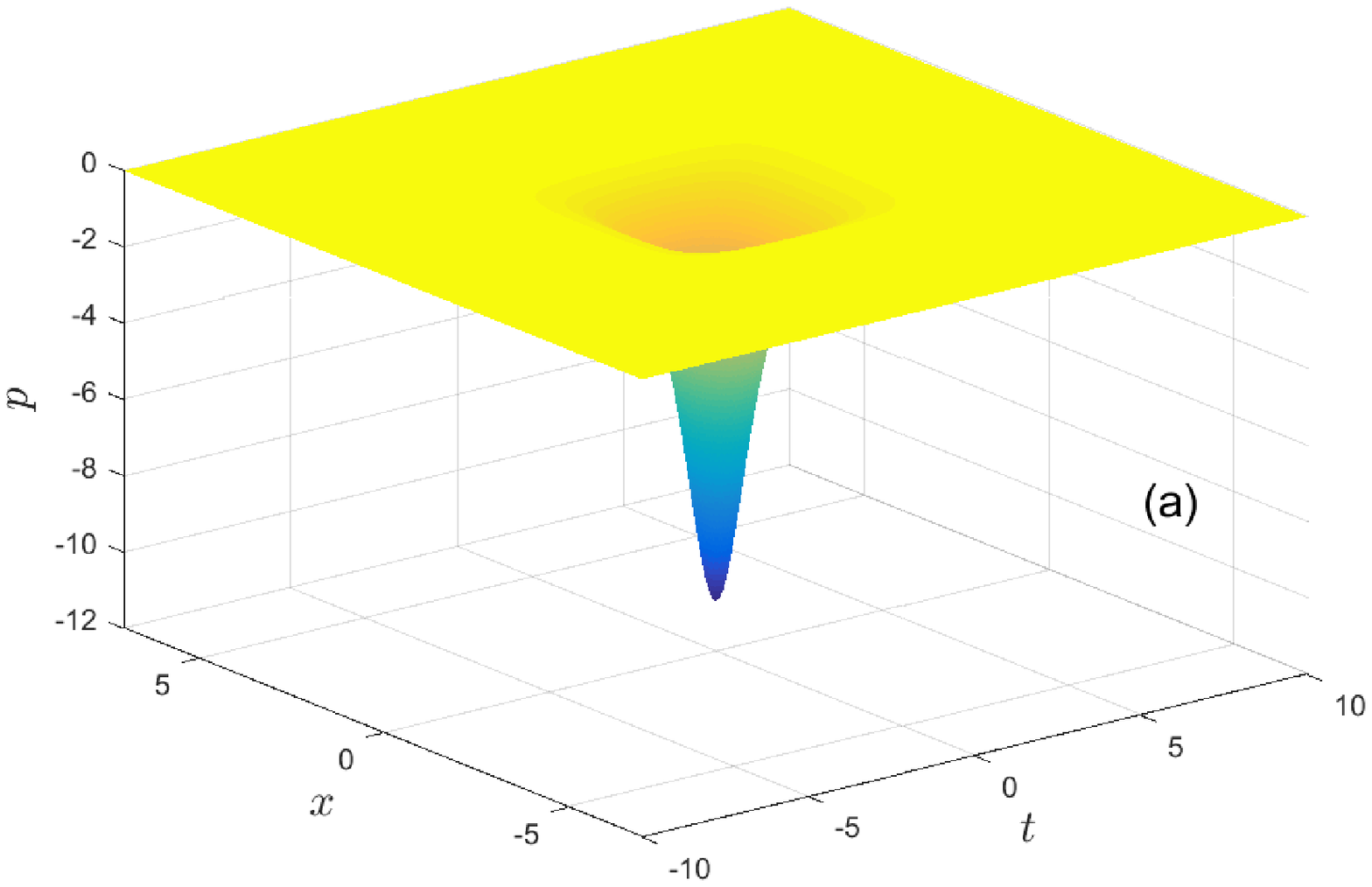}\epsfxsize=9cm\centerline{\hspace{-10.2cm}\epsfbox{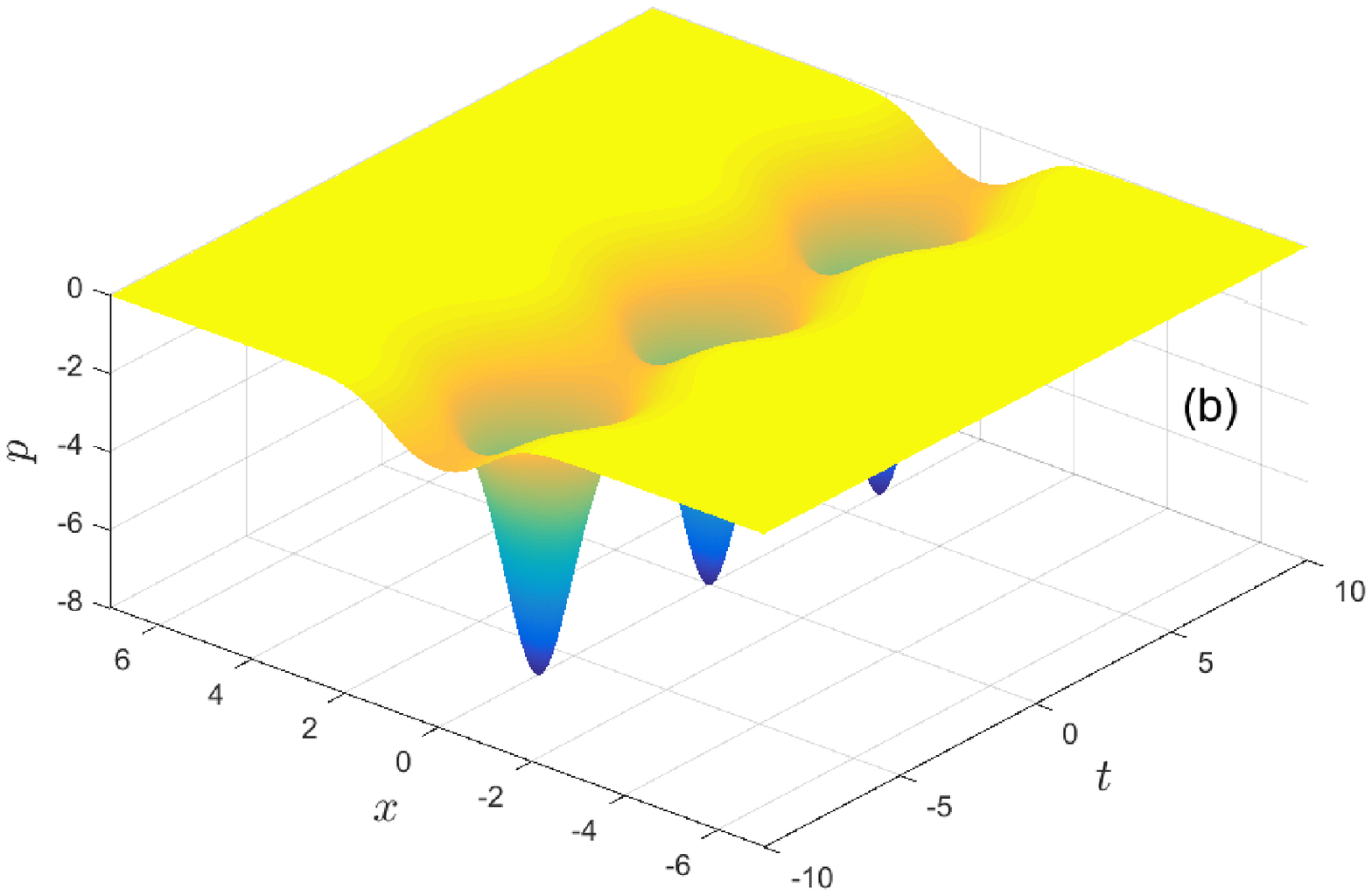}}}
\caption{Plots depict the pressure as a function of $x$ during time from $t=-10$ to $t=10$.
(a) For the kink-antikink solution  and (b) For the breather solution.\label{fig:pr}}
\end{figure}

\begin{figure}[h]
\epsfxsize=9cm\centerline{\hspace{8cm}\epsfbox{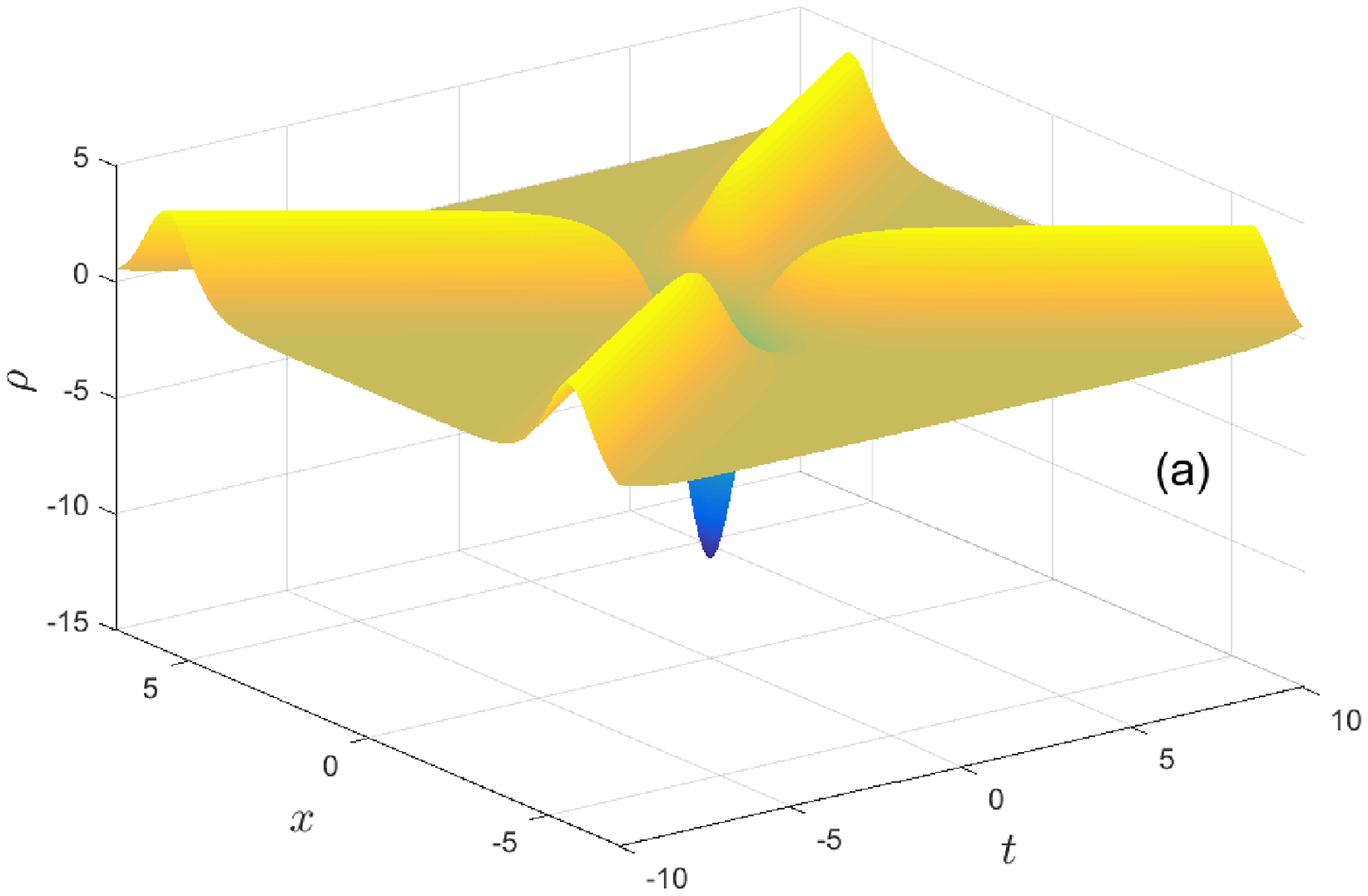}\epsfxsize=9cm\centerline{\hspace{-10.2cm}\epsfbox{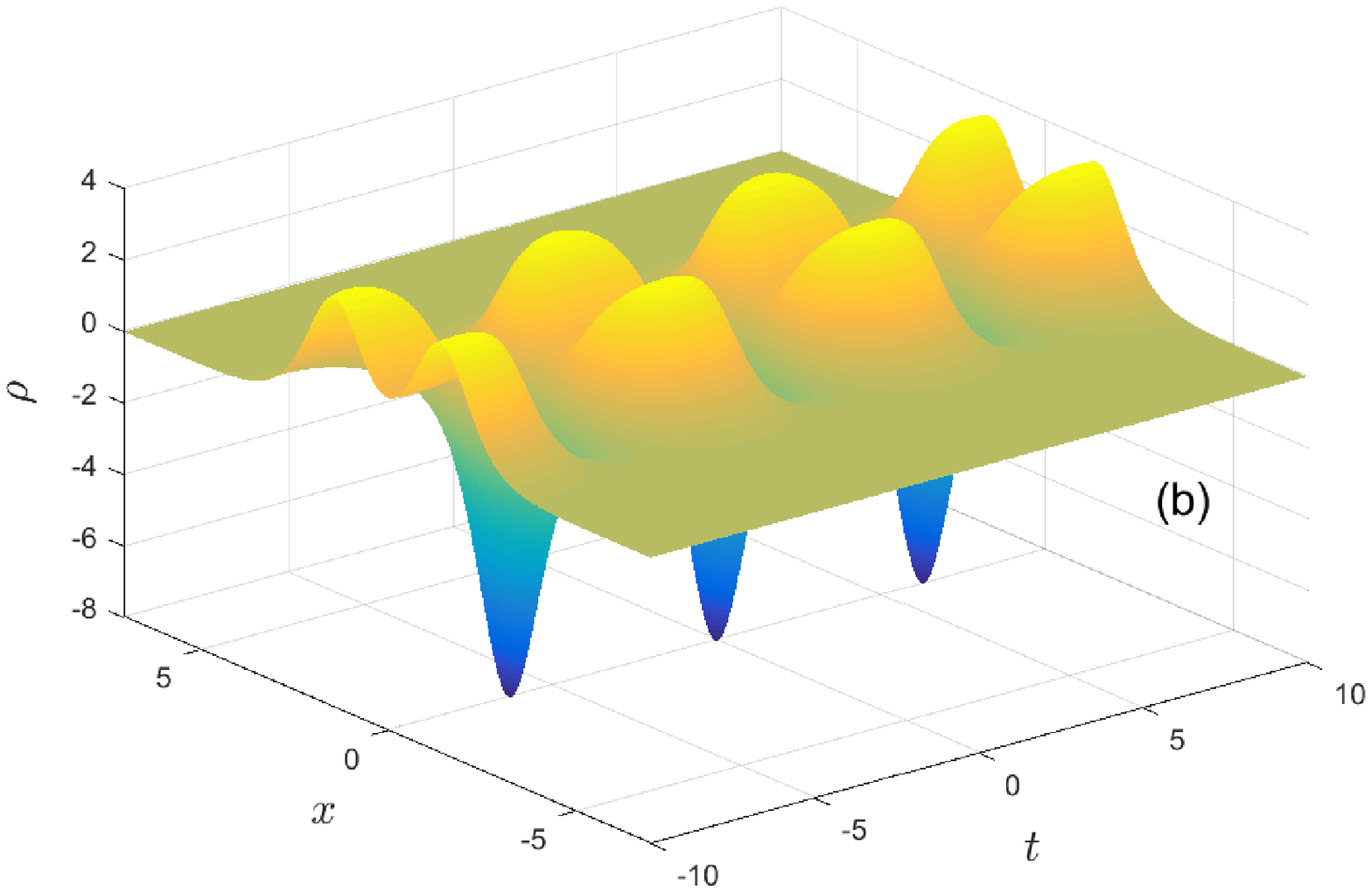}}}
\caption{Plots depict the energy density as a function of $x$ during the time from $t=-10$ to $t=10$;
(a) For the kink-antikink solution  and (b) For the breather solution.\label{fig:rho}}
\end{figure}

Instead of writing down the explicit forms of $\rho$ and $p$ as functions of $t$ and $x$ which are too lengthy, we prefer to plot the results for the typical value $w=1/2$ and $\gamma=2/\sqrt{3}$. Plots of $p$, $\rho$,  as functions of $x$ are shown in Figs.
\ref{fig:pr} and \ref{fig:rho}. $t\sim 0$ and $x\sim 0$ are the collision time and location, respectively. The existence of regions of high negative pressure can be attributed to the attracting force between solitons of opposite topological charges. Peaks in the density $\rho$ indicate the instantaneous position of the soliton. The behavior of kink-antikink scattering in the 
Sine-Gordon system is totally different from that of non-integrable models like the $\phi^4$, $\phi^6$ and $\phi^8$ systems \cite{hos,ba,mo1,mo2,mo3,mo4,mo5}.
%%%%%%%
%%%%%%%

%%%%%%%
%%%%%%%

%%%%%%%
%%%%%%%

%%%%%%%
%%%%%%%

%%%%%%%
%%%%%%%

%%%%%%%
%%%%%%%

%%%%%%%%%%%%%%%%%%%%%%%%%%%%%%%%%%%%%%%%%%%%%%%%%%%%%%%%%%%%%%%%%%%%%%%%%%%%%%%%
\section{Continuity Equation} \label{sec5}
%%%%%%%%%%%%%%%%%%%%%%%%%%%%%%%%%%%%%%%%%%%%%%%%%%%%%%%%%%%%%%%%%%%%%%%%%%%%%%%%

One of the basic equations of hydrodynamics is the continuity equation, which in its classical (non-relativistic) form is associated with the conservation of mass.
In its relativistic form, it takes the covariant form $\partial_{\mu}j^{\mu}=0$, governing the current $4$-vector $j^{\mu}$.  To this end, we start from the energy-momentum conversation $\partial_{\mu}T^{\mu\nu}=0$ with $\nu=0$. One is tempted to define $j^{\mu}\equiv T^{\mu0}$ as the relativistic current $4$-vector as mentioned in the introduction. However, one should note that $T^{\mu 0}$ is not necessarily a $4$-vector, but part of a tensor. The equation $\partial_{\mu}T^{\mu 0}=0$ is usually integrated over space, defining total energy as $E=\int T^{00} d^{N}x$, assuming vanishing flux density over the boundary, and reaching the conservation of total energy $\frac{dE}{dt}=0$. To achieve a consistent relation, one defines:
\begin{eqnarray}
j^{\mu}&\equiv&T^{\mu\nu}u_{\nu}=\left(\rho+p\right)u^{\mu}u^{\nu}u_{\nu}+u_{\nu}\eta^{\mu\nu}p,\nonumber\\
&=&\left[\rho+\left(1+\varepsilon\right)p\right]u^{\mu}.
\end{eqnarray}
This relation reduces to $j^{\mu}=\rho u^{\mu}$ for the time-like case $\varepsilon=-1$. However, $j^{\mu}$ is not conserved locally since:
\begin{equation}
\partial_{\mu}j^{\mu}=\left(\partial_{\mu} T^{\mu\nu}\right)u_{\nu}+T^{\mu\nu}\partial_{\mu}u_{\nu}=T^{\mu\nu}\partial_{\mu}u_{\nu}.
\end{equation}
The $4$-current $j^{\mu}$ is conserved, provided that $\partial_{\mu}u_{\nu}$ is purely anti-symmetric or $\partial_{[\mu}u_{\nu]}=0$.
For the bi-dimensional case $u^{\mu}=(u,v)$, this condition translates in to $\partial_{0}u=0$, $\partial_{1}v=0$ and $\partial_{1}u=-\partial_{0}v$.

%%%%%%%%%%%%%%%%%%%%%%%%%%%%%%%%%%%%%%%%%%%%%%%%%%%%%%%%%%%%%%%%%%%%%%%%%%%%%%%%
\section{concluding remarks}\label{con}
%%%%%%%%%%%%%%%%%%%%%%%%%%%%%%%%%%%%%%%%%%%%%%%%%%%%%%%%%%%%%%%%%%%%%%%%%%%%%%%

We attempted a fluid dynamical description of a relativistic scalar field and in particular the Sine-Gordon solitons. The main question we were to answer was the following: Is it possible to assign -in a consistent way- hydrodynamical, perfect fluid quantities pressure $p$, density $\rho$ and fluid velocity $u^{\alpha}$ to a single or double soliton solution? Is the fluid velocity $v$ necessary equal to the soliton velocity $w$? Is $\rho$ equal to $T^{00}$? We ended with the conclusion that the answers are not positive in a general sense. However, interesting relationships were found which may extend our knowledge about relativistic solitons. The procedure can straightforwardly be applied to non-perfect fluid models and solitons in dimensions higher than $1+1$, which is left for a future communication.

%%%%%%%%%%%%%%%%%%%%%%%%%%%%%%%%%%%%%%%%%%%%%%%%%%%%%%%%%%%%%%%%%%%%%%%%%%%%%%%%%%%

\acknowledgments{N. R acknowledges the support of Shahid Beheshti University Research Council. M. P acknowledges the support of Ferdowsi University of Mashhad.}

%%%%%%%%%%%%%%%%%%%%%%%%%%%%%%%%%%%%%%%%%%%%%%%%%%%%%%%%%%%%%%%%%%%%%%%%%%%%%%%%


\begin{thebibliography}{99}
%%%%%%%%%%%%%%%%%%%%%%%%%%%%%%%%%%%%%%%%%%%%%%%%%%%%%%%%%%%%%%%%%%%%%%%%%%%%%%%%
\bibitem{b} S. Weinberg, \textit{Gravitation and Cosmology: Principles and Applications of the General Theory of Relativity}, John Wiley and Sons, New York (1972).

\bibitem{a} S. Weinberg, S. Weinberg, ``Entropy generation and the survival of protogalaxies in an expanding universe,'' APJ {\bf 168}, 175 (1971).

\bibitem{c} G. L. Lamb, Jr, \textit{Elements of Soliton Theory}, John Wiley and Sons, New York (1980).

\bibitem{d} M. Guidry, \textit{Gauge Field Theories, an Introduction with
Applications}, Wiley, New York (1991).

\bibitem{Wein}
S. Weinberg, \textit{Cosmology}, Oxford University Press, New York (2008).


\bibitem{RD} R. D' Inverno, \textit{Introducing Einstein's  Relativity},
Oxford University Press, NewYork (1992).

\bibitem{e} N. Riazi and A. R. Gharaati, ``Dynamics of sine-Gordon solitons,'' Int. J. Theor. Phys. {\bf 37}, no. 3, 1081 (1998).

\bibitem{6}  O. M. Pimentel, F.D. Lora-Clavijo and G.A. Gonz\`{a}lez, ``The energy-momentum tensor for a dissipative fluid in general relativity," General Relativity and Gravitation {\bf 48}, 1 (2016).

\bibitem{7} C. Eckart, ``The thermodynamics of irreversible processes. III. Relativistic theory of the simple fluid," Phys. Rev. {\bf 58},  no. 10, 919 (1940).

\bibitem{8} C. W. Misner, K. S. Thorne and J. A. Wheeler, \textit{Gravitation}, W. H. Freeman and company San Fransisco, CA (1973).

 \bibitem{11} R. Jackiw, V.P. Nair, S.-Y. Pi, and A.P. Polychronakos, ``Perfect fluid theory and its extensions" J. Phys. A. (Math. and Gen.), {\bf 37}, no. 42, R327 (2004).
 
\bibitem{8a} G. M. Kremer, ``The Boltzmann equation in special and general relativity." In AIP Conference Proceedings, {\bf 1501}, no. 1, 160, American Institute of Physics, 2012. (2012).

\bibitem{Dh} T. Congy,  G. A. EL , G. Roberti, and A. Tovbis. ``Dispersive hydrodynamics of soliton condensates for the Korteweg-de Vries equation,'' arXiv preprint arXiv:2208.04472 (2022).

\bibitem{TB} T. Bonnemain, B. Doyon, and Gennady El. "Generalized hydrodynamics of the KdV soliton gas." Journal of Physics A: Mathematical and Theoretical {\bf 55}, no. 37, 374004 (2022).

\bibitem{EL} A. El Gennady, ``Soliton gas in integrable dispersive hydrodynamics,'' Journal of Statistical Mechanics: Theory and Experiment 2021, no. 11, 114001 (2021).

\bibitem{BD} B. Doyon, ``Lecture notes on generalised hydrodynamics,'' SciPost Physics Lecture Notes, 018 (2020).

\bibitem{zak} V. E. Zakharov, ``Turbulence in integrable systems,'' Studies in Applied Mathematics {\bf 122}, no. 3, 219 (2009).

\bibitem{10} A. Das, \textit{Integrable Models}, World Scientific Lecture Notes in Physics (1989).

\bibitem{man} N. Manton and P. Sutcliffe, \textit{Topological Solitons}, Cambridge University Press, Cambridge (2004).

\bibitem{Raj} R. Rajaraman, \textit{Solitons and Instantons: An Introduction to Solitons and Instantons in Quantum Field Theory}, North-Holland (1982).

\bibitem{Vach} T. Vachaspati, \textit{Kinks and Domain Walls An Introduction to Classical and Quantum Solitons}, Cambridge University Press, Cambridge (2006).

\bibitem{Ryder} L. H. Ryder, \textit{Quantum Field Theory, Second Edition}, Cambridge University Press, Cambridge (2001).

\bibitem{horita} R. Hirota, \textit{The Direct Method in Soliton Theory}, Cambridge University Press, Cambridge (2004).

\bibitem{9} T. Dauxois and M. Peyrard, \textit{Physics of Solitons}, Cambridge University Press, Cambridge (2006).

\bibitem{hos} S. Hoseinmardy and N. Riazi, ``Inelastic collision of kinks and antikinks in the $\phi^6$ system," {Int. J. Mod. Phys. A}, {\bf 25}, , no. 16, 3261 (2010).

\bibitem{ba}  R. H. Goodman and R. Haberman, "Kink-antikink collisions in the $\phi^4$ equation: The n-bounce resonance and the separatrix map," {SIAM J. Appl. Dyn. Sys.}, {\bf 4}, no. 4, 1195 (2005). 

\bibitem{mo1} A. Moradi Marjaneh, V. A. Gani, D. Saadatmand, S. V. Dmitriev, and K. Javidan, ``Multi-kink collisions in the $\phi^6$ model," Journal of High Energy Physics 2017, no. 7, 1 (2017).

\bibitem{mo2}  A. Askari, A. Moradi Marjaneh, Z. G. Rakhmatullina, M. Ebrahimi-Loushab, D. Saadatmand, Vakhid A. Gani, P. G. Kevrekidis, S. V. Dmitriev, ``Collision of $\phi^4$ kinks free of the Peierls–Nabarro barrier in the regime of strong discreteness,'' Chaos, Solitons and Fractals {\bf 138}, 109854 (2020).  

\bibitem{mo3} V. A. Gani, A. Moradi Marjaneh and Kurosh Javidan, ``Exotic final states in the $\phi^8$ Multi kink collisions ,'' Eur. Phys. J. C {\bf 81}, 1124 (2021).

\bibitem{mo4} D. Saadatmand and A. Moradi Marjaneh, ``Scattering of the asymmetric $\phi^6$ kinks from a PT -symmetric perturbation: creating multiple kink–antikink pairs from phonons ,''  Eur. Phys. J. B {\bf 95}, 144 (2022).

\bibitem{mo5} A. Moradi Marjaneh, F. C. Simas, D. Bazeia, ``Collisions of kinks in deformed $\phi^4$ and $\phi^6$ models, '' 
Chaos, Solitons and Fractals {\bf 164}, 112723 (2022).

\end{thebibliography}
\end{document}